\documentstyle[12pt,aasms4]{article}

\begin{document}
\onecolumn

\title{
Classifications of the Host Galaxies of Supernovae, Set II
}

\author{Sidney van den Bergh}
\affil{Dominion Astrophysical Observatory, Herzberg Institute of Astrophysics,
National Research Council, 5071 West Saanich Road, Victoria, British
Columbia, V9E 2E7, Canada (sidney.vandenbergh@nrc.gc.ca)
}

\centerline{and}

\author{Weidong Li and Alexei V. Filippenko}
\affil{
Department of Astronomy, 601 Campbell Hall, University of
California, Berkeley, CA 94720-3411 (wli@astro.berkeley.edu,
alex@astro.berkeley.edu)
}

\newpage

\begin{abstract}

   Classifications on the DDO system are given for an additional 231 host
galaxies of supernovae that have been discovered during the course of the Lick
Observatory Supernova Search with the Katzman Automatic Imaging Telescope (KAIT).
This brings the total number of hosts of supernovae discovered (or independently
rediscovered) by KAIT, which have so far been classified on a homogeneous system,
to 408.  The probability that SNe~Ia and SNe~II have a different distribution of
host galaxy Hubble types is found to be 99.7\%. A significant difference is also
found between the distributions of the host galaxies of SNe~Ia and of SNe~Ibc
(defined here to include SNe~Ib, Ib/c, and Ic).  However, no significant difference
is detected between the frequency distributions of the host galaxies of SNe~II and
SNe~IIn.  This suggests that SNe~IIn are generally not SNe~Ia embedded in circumstellar
material that are masquerading as SNe~II. Furthermore, no significant difference
is found between the distribution of the Hubble types of the hosts of SNe~Ibc 
and of SNe~II. Additionally, SNe~II-P and SNe~II-L are found to occur among similar
stellar populations. The ratio of the number of SNe~Ia-pec to normal SNe~Ia appears
to be higher in early-type galaxies than it is in galaxies of later morphological types.
This suggests that the ancestors of SNe~Ia-pec may differ systematically in age 
or composition from the progenitors of normal SNe~Ia. Unexpectedly, five SNe of
Types Ib/c, II, and IIn (all of which are thought to have massive progenitors)
are found in host galaxies that are nominally classified as types E and S0. However,
in each case the galaxy classification is uncertain, or newly inspected images show
evidence suggesting a later classification. Among these five objects NGC 3720, the host
galaxy of SN 2002at, was apparently misidentified in the Carnegie Atlas of Galaxies.  

\end{abstract}

 
\section{Introduction}

   Archaeologists derive much of their knowledge of ancient civilizations from
digging in ancient cemeteries. By the same token astronomers can learn a great
deal about the evolutionary history of galaxies from the frequency, and
distribution of, supernova (SN) outbursts which signal the occurrence of
stellar deaths.

   Until quite recently the discovery of a supernova was a rather rare and
haphazard event. However, the advent of modern systematic and automated
searches now enables us to study the occurrence of supernovae (SNe) in a much
more systematic fashion. The largest such automated search program is currently
the Lick Observatory Supernova Search (LOSS) with the 0.75-m Katzman Automatic
Imaging Telescope (KAIT). LOSS, which started in 1997 and recently expanded to
the Lick Observatory and Tenagra Observatory Supernova Searches (LOTOSS;
Schwartz et al. 2000), has been described by Treffers et al. (1997), Li et
al. (2000), Filippenko et al. (2001), and Filippenko (2003). In order to derive
the full benefit from LOSS, one needs to know both the spectroscopic types of
all of these automatically discovered SNe and the morphological classifications
of the SN host galaxies. By combining the new morphological classifications of
231 host galaxies with those of 177 hosts in a previous paper, we now have
available morphological types for the host galaxies for a total of 408
SNe that were discovered (or independently rediscovered) during the course of
the LOSS/LOTOSS surveys. This large and homogeneous data set will be discussed
below.

\section{New Morphological Classifications}

   The present investigation represents a continuation of the work of van den
Bergh, Li, \& Filippenko (2002, hereafter Paper I), in which the Hubble types
of 177 SNe discovered during the course of LOSS were classified on the DDO
System (van den Bergh 1960a,b,c). In the DDO system galaxies are assigned a
Hubble type [E0...E7, S0, Sa, Sb, Sc, Ir], a luminosity class [I...V], and to a
form family [S, S(B) or SB]. An ``n" denotes smooth-armed spirals, a ``*"
denotes patchy-armed spirals, a ``t" denotes tidally distorted objects, and a
``:" is used for uncertain values.

   New classifications of the host galaxies of SNe that were discovered (or
rediscovered) during the course of LOSS/LOTOSS are listed in Table I. For
brevity, in the table N = NGC, I = IC, U = UGC, and M = MCG. In addition to the KAIT CCD
images themselves, we also had available either the red images of the Palomar
Sky Survey (POSS), or the SERC-J survey in blue-green light. For southern
galaxies red IIIaF + RG610 images were also available from the UK Schmidt
Equatorial Red Survey or from the UK Schmidt Second Epoch red survey. For
additional information on the observational database the reader is referred to
Paper I. For some galaxies only images in red light were available. This posed
no problem for elliptical galaxies, but required some ``mental extrapolation"
for the spirals that had no $B$ images available. All morphological
classifications were made by one of the authors (SvdB). Because of the limited
dynamic range of the images it was often not possible to distinguish with
confidence between E and S0 galaxies. Although the majority of galaxies with
redshift $z \approx 0.0$ fit comfortably within the Hubble classification
system, there are some that do not. No attempt was made to shoe-horn peculiar
objects into the Hubble scheme.

\section{Classifications of Galaxy Morphology and Supernova Type}

\subsection{Luminosity Classifications}

   Table 1 contains classifications on the DDO system for 44 galaxies that had
also been classified previously by van den Bergh (1960c). One of these objects
(NGC 4653) was apparently misidentified in that paper and has been excluded
from the comparisons given below. For the remaining 24 galaxies that had
luminosity classifications the mean difference (in the sense old minus new) was
$-0.02 \pm 0.09$ luminosity classes. The root-mean-square dispersion in the
difference between the old and the new classifications was 0.43 luminosity
classes, so that the intrinsic dispersion of a single luminosity classification
on the DDO system is $\sim$0.3 luminosity class.  This close agreement is
particularly pleasing because our comparison includes objects with uncertain
luminosity classifications (that had been marked ``:").

\subsection{Classifications of Hubble Types}

   Of the host galaxies of SNe that are listed in Table 1, 36 were also
assigned Hubble types by van den Bergh (1960c). Thirty galaxies (83\%) were
independently assigned the same Hubble type in 1960 and in 2003, 4 (11\%) were
classified half a Hubble class later in the 2003 investigation, and 2 (6\%)
were assigned half a Hubble class earlier in the 2003 classification. For no
object did the 1960 and the 2003 classifications differ by 1.0 or more Hubble
classes. It is remarkable (and satisfying!) to see that morphological
classifications made independently (but by the same person) 
with a time separation of over four decades
exhibit no measurable systematic differences.

\subsection{Spectral Classifications}

   The classifications of SN spectra that are given in Table 1 were drawn from
the IAU Circulars. Supernovae of Type Ia were divided into ``normal" and
``peculiar" categories on the basis of careful inspection of the spectroscopic
information in the IAU Circulars: objects that showed the strong Si~II
$\lambda$5970 feature or Ti~II absorption lines near 4200~\AA\ (which are
evidence for a subluminous SN 1991bg-like event; Filippenko 1997), or weak
Si~II $\lambda$6350 absorption or strong Fe~III absorption (which indicates a
possibly overluminous, SN 1991T-like event; Filippenko 1997) were classified as
``peculiar" SNe~Ia.

\section{Discussion} 

\subsection{Possible Massive Supernovae in E and S0 Galaxies}

   It is generally believed (e.g., van den Bergh \& Tammann 1991) that only SNe
of Type Ia occur among the old stellar populations that inhabit elliptical
galaxies. The greatly enlarged modern database of accurate DDO morphological
classifications and uniform SN classifications provided in the present paper,
and in Paper I, allows us to test this hypothesis.

   Table 2 lists the five SNe, which are not of Type Ia, that have been
discovered in host galaxies that appear to be of types E or S0.
It should be emphasized that late-type
galaxies of unusually high surface brightness may be misclassified
as being of early type. A good example of this
effect is provided by the high surface brightness face-on
late-type spiral NGC 3928 = Markarian 190 (van den Bergh 1980)
which was misclassified as an E0 in the Second Reference
Catalogue (de Vaucouleurs, de Vaucouleurs, \& Corwin 1976). It would be of
interest to obtain large-scale images of the galaxies that are listed in Table
2 to see if any of these galaxies were misclassified, or show evidence for a
recent merger with an object of later type that might have contained massive
stars. Alternatively, a few early-type galaxies with late-type companions might
have picked up some gas during tidal encounters.
In this connection it is noted that Sandage \& Bedke (1994) refer to
very subtle dust patches in NGC 2768.  Massive stars and SN
progenitors might, of course, also have formed recently in the gas that was so
acquired. Finally, some of the SNe~II that appear to be located in E or S0
galaxies might actually be situated in dim late-type companions to these
objects.

   On the basis of the data listed in Table 2 it appears that not more than
$\sim$3\% of all SNe with massive progenitors occur in E and S0 galaxies.
However, note that some of our E/S0 classifications might be erroneous because
the SN host galaxy was a late-type object with an unusually high surface
brightness. Such a high surface brightness might, for example, result from an
intense burst of star formation triggered by a rather recent tidal encounter.
As listed in the notes of Table 2, out of the 5 galaxies mentioned, two
(UGC 2836 and NGC 3720) are possible Sa galaxies, and two (IC 2461 and IC 3203)
are viewed edge-on and have somewhat uncertain classifications (note,
for example, NED\footnote{NED is the NASA/IPAC Extragalactic Database,
http://nedwww.ipac.caltech.edu.} lists the two galaxies as Sb and Sbc,
respectively). Even NGC 2768 might not be a convincing early-type galaxy, being
classified as E3/Sa; moreover, it is a LINER and may host a nuclear starburst. 
It is thus inconclusive whether {\it any} SNe with massive progenitors have 
been discovered in normal early-type (E/S0) galaxies from this study. 

\subsection{Relative Supernova Numbers in Galaxies of Differing Types}

   The number of SNe of various types in Table 1 is listed in Table 3. In Table
4 these new data are combined with those that we had previously published in
Paper I. Inspection of these data reveals the following trends:

\begin{enumerate}

\item{A Kolmogorov-Smirnov test shows that the distribution of 117 SNe~Ia 
   and 119 SNe~II (excluding SNe~IIn) over host galaxy type is different at 
   99.7\% confidence. The observed difference is in the expected sense, 
   with the SNe~II (which are believed to have massive core-collapse 
   progenitors) being most common in late-type galaxies of 
   Hubble types Sb--Sc, whereas SNe~Ia are observed to occur in 
   hosts of all Hubble types. A small number of possible cases in 
   which SNe~II appear to have occurred in environments that seem
   to be overwhelmingly composed of old stars is given in Table 2. 
   Note, however, that such objects account for only $\sim$3\% of our
   total sample of core-collapse SNe (i.e., SNe~II, Ib, Ib/c, Ic).}

\item{A comparison between the distributions of 119 SNe~II and of 
   15.5 SNe~IIn shows no hint of any difference in the distribution
   of host types for these two kinds of SNe, but this result
   suffers from small-number statistics (an object classified as S0/Sa
   was counted as 0.5 S0 and 0.5 Sa). In this 
   connection it is noted that Hamuy et al. (2003) have recently 
   speculated that some SNe~IIn might be SNe~Ia that are interacting 
   with circumstellar material. If our conclusion is correct, most 
   SNe~IIn are probably {\it not} SNe~Ia masquerading as SNe~II.
   Only if SNe~IIn come from the subset of SNe~Ia having the most massive
   possible progenitors could the two observed distributions be consistent.}

\item{A comparison between the distributions of normal and peculiar SNe~Ia 
   hints at the possibility that SNe~Ia-pec might occur preferentially
   in early-type hosts. A Kolmogorov-Smirnov test shows that there
   is only a 7\% probability that they are
   drawn from the same parent population of host galaxies. Since
   some normal SNe~Ia and SNe~Ia-pec occurred in S0 galaxies, which lie
   outside the linear E-Sa-Sab-Sb-Sbc-Sc-Ir scheme, the sample can
   be enlarged by dividing galaxies into (1) elliptical and S0
   galaxies, and (2) spiral and irregular galaxies. Such a division
   has the advantage that it allows us to include in the statistics 
   spirals classified as ``S" but that could not 
   be assigned to types ``Sa," ``Sb," or ``Sc." 
   Table 5 shows a compilation of the frequency  
   distribution of normal SNe~Ia and SNe~Ia-pec in E + S0 galaxies and
   in S + Ir galaxies. For the data in this table, $\chi^2 = 4.7$.
   A $\chi^2$ test with one degree of freedom therefore yields 
   only a 3\% chance that SNe~Ia-pec have the same probability 
   distribution over Hubble types as do normal SNe~Ia. }

\item{There is not even a hint of a difference in the distribution of
   host-galaxy Hubble types between 45.5 SNe~Ibc (defined here to include 
   SNe~Ib, Ib/c, and Ic) and 119 SNe~II.
   Moreover, although the currently available data sample is rather
   small, there is no evidence that SNe~Ib, SNe~Ib/c, and SNe~Ic have
   different distributions over host-galaxy types.}

\item{A Kolmogorov-Smirnov test of the data for 117 SNe~Ia and 46.5 
   SNe~Ibc shows that there is only a 0.1\% probability that the 
   host galaxies of these two types of objects are drawn from the 
   same distribution of Hubble types.}

\end{enumerate}

   These results are consistent with the widely held suspicion that SNe~Ib, 
SNe~Ib/c, SNe~Ic, and SNe~II all have massive progenitors, whereas 
those of SNe~Ia do not.

\subsection{Peculiar Type Ia Supernovae}

   The data in Table 1 show that the majority of the 16 SNe~Ia with spectra
resembling that of the peculiar, subluminous SN 1991bg occurred in early-type
galaxies, whereas all three of those with spectra like the peculiar, possibly
overluminous SN 1991T erupted in spirals of types Sb or Sc. A
Kolmogorov-Smirnov test shows that there is only a 4\% probability that objects
like SN 1991bg and those resembling SN 1991T are drawn from the same
distribution of host galaxy classification types. Our results suggest that
luminous SNe~Ia resembling SN 1991T probably had younger (or more massive)
progenitors than do those of the SN 1991bg variety.  The present data confirm
and strengthen a similar conclusion by Howell (2001).

\subsection{Supernovae of Types II-L and II-P}

   Ever since Minkowski (1941), it has been customary to classify SNe primarily
according to their spectra.  However, it is also possible to classify them from
their light-curve morphology. This enables one to assign SNe~II to ``L"
(linear) and ``P" (plateau) subtypes (e.g., Doggett \& Branch 1985).  Schlegel
(1996) and Filippenko (1997) have suggested that SNe~II-P and SNe~II-L can be
spectroscopically distinguished on the basis of the presence or absence of
H$\alpha$ absorption. However, this speculation still needs to be confirmed
with additional data.

   A recent listing of SNe~II that have been assigned to the linear and plateau
subtypes is given in the Asiago Supernova Catalog
(http://merlino.pd.astro.it/$\sim$supern/). For the majority of these objects,
high quality and homogeneous morphological classifications of the host
galaxies, based on inspection of plates obtained with large reflectors, are
given by Sandage \& Tammann (1981). These data show a very similar distribution
of SNe~II-L and of SNe~II-P over Hubble types. It is found that 17 out of 25
(68\%) of all SNe~II-P occur in galaxies of type Sc. This does not differ
significantly from the distribution of SNe~II-L, for which 8 out of 11 (73\%)
are located in spirals of type Sc.  Luminosity classes are available for only
34 of the host galaxies of SNe assigned to types II-P and II-L.  For this small
sample there is a hint that SNe~II-L might occur in more luminous (metal-rich?)
galaxies than do SNe~II-P. However, a Kolmogorov-Smirnov test of the data shows
that this possible difference does not have a respectable level of statistical
significance. It is therefore concluded that the currently available database
does not provide evidence for a significant difference between the host
populations of SNe~II-P and SNe~II-L.

\section{Summary}

   Morphological types on the DDO system are now available for a homogeneous
sample of the host galaxies of 408 SNe that were discovered (or rediscovered
independently) during the course of the LOSS/LOTOSS surveys. Based on these
data, we come to the following conclusions.

\begin{enumerate}

\item{The present (2003) classifications of Hubble types and luminosity
  classes by van den Bergh are, to within statistical errors, found 
  to be on exactly the same system as his DDO classifications that
  were made in 1959 and 1960.}

\item{Five SNe of Types Ib/c, II, and IIn (which are thought to have 
  massive progenitors) are found to have occurred in host galaxies 
 nominally classified as types E or S0. However, in each case the galaxy
  classification is uncertain, or newly inspected images show evidence
  suggesting a later classification. Thus, we find no clear evidence of
  SNe with massive progenitors occurring in E or S0 galaxies. It would 
  clearly be important to obtain new, large-scale images of the five
  early-type galaxies to improve the classifications and to search for 
  possible evidence of young population components.}

\item{The distribution of SNe~Ia and SNe~Ia-pec over host-galaxy Hubble
  types is found to differ from that for SNe~II + SNe~IIn at 99.7\% 
  confidence.}

\item{No statistically significant differences are found between the
     distributions over Hubble type of the host galaxies of SNe~II
     and SNe~IIn. Hence, it appears unlikely that most SNe~IIn are
     actually SNe~Ia embedded in circumstellar material.}

\item{Compared with normal SNe~Ia, those objects classified as SNe~Ia-pec
     are found to be more common in early-type galaxies than they
     are in spirals of later type. A Kolmogorov-Smirnov test shows
     only a 1.6\% probability that the 117 SNe~Ia and the 29.5 
     SNe~Ia-pec in Table 4 are drawn from the same parent population
     of Hubble types. This suggests that the progenitors of normal
     SNe~Ia may differ in age or metallicity from the ancestors 
     of SNe~Ia-pec.}

\item{Among SNe~Ia-pec, those that resemble SN 1991T appear to be
     younger (or more massive) than those like SN 1991bg --- i.e.,
     all three SN 1991T-like objects occurred in spirals of types
     Sb to Sc, whereas 14 out of 16 objects resembling SN 1991bg were
     hosted by galaxies of type E, S0, or Sa. A Kolmogorov-Smirnov 
     test shows only a 1.4\% probability that the three SN 1991T-like 
     and the sixteen SN 1991bg-like objects are drawn from the same 
     population of host galaxy Hubble types.}

\item{There is only a 0.1\% probability that the SNe~Ia and SNe~Ibc 
   (defined here to include SNe~Ib, Ib/c, and Ic) in the present sample 
   are drawn from the same parent population. SNe~Ia are found to occur 
   in all kinds of galaxies, whereas the (presumably more massive) 
   progenitors of SNe~Ibc mostly occur in spirals of types Sb--Sc.}

\item{No significant differences are found between the distributions 
  over Hubble type of the small numbers of SNe~Ib, SNe~Ib/c, and 
  SNe~Ic contained in the present sample.}

\item{SNe~II-P and SNe~II-L appear to occur in galaxies having similar 
  stellar populations.}

\end{enumerate}

   All of the data discussed above refer to the relative rates of SNe of
different types. The implications of the present morphological classifications
for absolute frequency determinations of SNe of different types will be the
subject of a future investigation.

\acknowledgments

    We are indebted the the referee (Harold G. Corwin, Jr.) for
a number of helpful suggestions and for providing useful
information on various galaxies. We are also grateful 
to Allan Sandage for information on the classification of
NGC 3719 and NGC 3720, to David Branch for consultations on SN spectra,
to Ryan Chornock for helpful discussions, and to Avishay Gal-Yam for
valuable information that affected our discussion of the core-collapse 
SNe that may have occurred in early-type host galaxies.
The work of A.V.F.'s group at U. C. Berkeley is supported by National Science
Foundation grant AST-0307894, as well as by the Sylvia and Jim Katzman
Foundation. KAIT was made possible by generous donations from Sun Microsystems,
Inc., the Hewlett-Packard Company, AutoScope Corporation, Lick Observatory, the
National Science Foundation, the University of California, and the Katzman
Foundation.

\newpage

\renewcommand{\arraystretch}{0.75}

\begin{deluxetable}{llllll}
\tablecaption{Classifications of SN Host Galaxies}
\label{1}
\tablehead{
\colhead{SN} & \colhead{Galaxy} & \colhead{DDO Type} &
\colhead{SN Type} & \colhead{Velocity (km/s)} & \colhead{Remarks}
}
\startdata 
1998dl &   N1084     & Sc II:           &   II   &    1406 & \\
1998dn &   N 337A    & Sc pec           &   II   &    1094& \\
1998ec &   U3576     & SBb              &   Ia   &    5966    & \\
1998ey &   N7080     & S(B)bc I         &   Ic-pec  &    4839& \\
1999D &    N3690     & S + Pec          &   II   &    3121  & \\
1999aa &   N2595     & S(B?)bc pec II:  &   Ia-pec (91T)  &    4330  & \\
1999an &   I 755     & Sb: III-IV       &   II   &    1507  &2 \\
1999di &   N 776     & S(B)bc I-II      &   Ib   &    4921& \\
1999dn &   N7714     & Stt              &   Ib/c  &    2798  &3 \\
1999eh &   N2770     & Sbc: III-IV?     &   Ib   &    1947& \\
1999el &   N6951     & S(B)b I-II       &   IIn  &    1424& \\
1999ev &   N4274     & Sab II-III       &   II   &     930& \\
1999gi &   N3184     & Sc I             &   II   &     592& \\
1999gk &   N4653     & Sc II:           &   II   &    2626& \\
1999gn &   N4303     & Sc I             &   II   &    1566& \\
2000B &    N2320     & S0/E3            &   Ia   &    5944 & \\
2000C &    N2415     & ?                &   Ic   &    3784  &4 \\
2000E &    N6951     & SBb I-II         &   Ia   &    1424& \\
2000K &    M+09-19-191&E2               &   Ia   &   16000   & \\
2000L &    U5520     & Sb II-III:       &   II   &    3315 & \\
2000M &    N6389     & Sbc II           &   II   &    3119& \\
2000bk &   N4520     & E3/Sa            &   Ia   &    7628& \\
2000ce &   U4195     & SBb II           &   Ia   &    4888& \\
2000cr &   N5395     & Sbt I?           &   Ic   &    3491& \\
2000cs &   M+07-34-15& Sab              &   II   &   10532& \\
2000cz &   I1535     & Sb III-IV        &   II   &    5231  & \\
2000db &   N3949     & Sbc III:         &   II   &     800& \\
2000ds &   N2768     & E3/Sa            &   Ib/c  &    1373 & \\
\enddata
\tablenum{1}
Note:---Remarks: (1) merger; (2) edge-on; (3) tides; (4) compact, high surface 
brightness; (5) ring; (6) in a compact cluster; (7) galaxy too faint to classify;
(8) SN 2003A is probably in the Sb galaxy U5904, even though it is closer to
the E4 galaxy U5907; (9) SN 2003H occurs halfway between two Sc galaxies,
either of which could be the host galaxy; (10) possibly two nuclei.
\end{deluxetable}

\newpage

\begin{deluxetable}{llllll}
\tablecaption{(continued)}
\tablehead{
\colhead{SN} & \colhead{Galaxy} & \colhead{DDO Type} &
\colhead{SN Type} & \colhead{Velocity (km/s)} & \colhead{Remarks}
}
\startdata
2000ew &   N3810     & Sc I-II          &   Ic   &     993& \\
2000ez &   N3995     & Sc? t            &   II   &    3254& \\
2000fn &   N2526     & Sa               &   Ib   &    4603& \\
2000fs &   N1218     & E2               &   Ia   &    8590& \\
2000ft &   N7469     & Sb: pec          &    ?   &    4892& \\
2000fu &   M-03-38-21& S(B?)b:          &    ?   &    3265& \\
2000fv &   M-04-35-11& SBab             &    ?   &    2738& \\
2001H &    M-01-10-19& Sb II            &   II   &    5248& \\
2001K &    I 677     & Sab:             &   II   &    3249& \\
2001T &    M-02-37-6 & Sb II-III:       &   II   &    4167& \\
2001V &    N3987     & Sb I?            &   Ia-pec (91T)  &    4502   &2 \\
2001X &    N5921     & SBb I            &   IIp  &    1480& \\
2001aa &   U10888    & SBb I-II         &   II   &    6110& \\
2001ad &   N6373     & Sc pec           &   IIb  &    3320& \\
2001bg &   N2608     & Sc II:           &   Ia   &    2135& \\
2001bq &   N5534     & Sb: t            &   II   &    2633  & \\
2001br &   U11260    & SBb II-III:      &   Ia   &    6184   & \\
2001cf &   U7020     & Sbc I-II         &   IIb  &    6132& \\
2001cm &   N5965     & Sb               &   II   &    3412    &2 \\
2001dc &   N5777     & Sb               &   II-pec  &    2145    &2 \\
2001de &   U12089    & Sbc              &   Ia-pec (91bg)  &    9313& \\
2001dn &   N 662     & Ir?              &   Ia   &    5654& \\
2001dp &   N3953     & S(B)bc I         &   Ia   &    1052& \\
2001dr &   N4932     & S(B?)c II        &   II   &    7088    & \\
2001dw &   N1168     & S(B)b II         &   Ia   &    7627& \\
2001dz &   U 471     & Sc II:           &   II   &   14810& \\
2001eb &   N1589     & Sab              &   Ia   &    3793    &2 \\
2001ed &   N 706     & Sc II-III:       &   Ia   &    4980& \\
2001ee &   N2347     & Sb III           &   II   &    4421& \\
2001eg &   U3885     & Sbc III:         &   Ia   &    3809& \\
2001ej &   U3829     & St + S?          &   Ib   &    4031   &    1? \\
2001eo &   U3963     & Sb III-IV:       &   Ia   &     ...   &    2 \\
2001ew &     ...     & E2/Sa            &   Ia   &     ...& \\
2001ex &   U3595     & SBbc II          &   Ia   &    7908& \\
2001fe &   U5129     & Sa               &   Ia   &    4059& \\
2001fv &   N3512     & Sc II-III        &   II   &    1376 & \\
2001fz &   N2280     & Sb I-II          &   II   &    1906& \\
2001gd &   N5033     & Sbc II           &   IIb  &     875& \\
2001hg &   N4162     & Sc II            &   II   &    2569& \\
2001ib &   N7242     & E2               &   Ia-pec (91bg)  &    5790& \\
2002an &   N2575     & Sbc III          &   II   &    3870& \\
\enddata
\tablenum{1}
\end{deluxetable}

\newpage
\begin{deluxetable}{llllll}
\tablecaption{(continued)}
\tablehead{
\colhead{SN} & \colhead{Galaxy} & \colhead{DDO Type} &
\colhead{SN Type} & \colhead{Velocity (km/s)} & \colhead{Remarks}
}
\startdata
2002ap &   N 628     & Sc I             &   Ib/c-pec &     657& \\
2002at &   N3720     & E1               &   II   &    5985   & \\
2002au &   U5100     & SBbc II-III:     &   Ia   &    5514 & \\
2002av &   ESO 489-G7& E0 pec           &   Ia   &     ...  &     5? \\
2002aw &     ...     & Sb/S0            &   Ia   &     ...  &     2 \\
2002bf &   CGCG 266-031& Sa               &   Ia   &    7254& \\
2002bg &   M+02-38-31& Sbc II           &   Ia   &   12814& \\
2002bh &   U5286     & Sc II-III        &   II   &    5198 & \\
2002bi &   U8527     & S                &   Ia   &    6986   &    2 \\
2002bj &   N1821     & Sc:              &   IIn  &    3608& \\
2002bl &   U5499     & Sb III:          &   Ib/c-pec &    4753& \\
2002bm &   M-01-32-19& SBbc I           &   Ic   &    5462& \\
2002bo &   N3190     & Sa III           &   Ia   &    1271    &   2 \\
2002bp &   U6332     & Sa:              &   ?    &    6227    &   5 \\
2002bs &   I4221     & Sc               &   Ia   &    2895& \\
2002bt &   U8584     & St+E+S           &   Ia   &   17859    &   1 \\
2002bu &   N4242     & S/Ir IV          &   IIn  &     517& \\
2002bv &   U4042     & SBb II-III       &   IIn  &    8292  & \\
2002bw &     ...     & S?               &   Ia   &    5197& \\
2002bx &   I2461     & S0               &   II   &    2260    &   2 \\
2002bz &   M+05-34-33& E1               &   Ia   &   11090  & \\
2002ca &   U8521     & SBb: II-III      &   II   &    3277& \\
2002cc &     ...     & Sab              &   Ia   &   19950& \\
2002cd &   N6916     & Sc:              &   Ia   &    3101& \\
2002ce &   N2604     & SB IV?           &   II   &    2094& \\
2002cf &   N4786     & E2               &   Ia-pec (91bg) &    4647  & \\
2002cg &   U10415    & Sc I-II:         &   Ic   &    9546& \\
2002ci &   U10301    & S(B?) III:       &   Ia   &    6663& \\
2002cj &   ESO 582-G5& Sb               &   Ic   &    6758& \\
2002ck &   U10030    & SBb III          &   Ia   &    8953& \\
2002cp &   N3074     & Sc I-II          &   Ib/c  &    5144& \\
2002cr &   N5468     & Sc II            &   Ia   &    2845 & \\
2002cs &   N6702     & E2               &   Ia   &    4728& \\
2002cu &   N6575     & E2               &   Ia-pec (91bg)  &    6992& \\
2002cv &   N3190     & Sab t            &   Ia   &    1271    &   2 \\
2002cw &   N6700     & SBbc II-III      &   Ib   &    4588& \\
2002cy &   N1762     & Sab:             &   ?    &    4753& \\
2002db &   N5683     & Sa:              &   Ia   &   10859& \\
2002de &   N6104     & SB IV            &   Ia   &    8428& \\
2002df &   M-01-53-6 & S(B)b II:        &   Ia   &      ... & \\
2002di &   M+05-40-2 & Sa:              &   Ia-pec (91bg)  &   10910& \\
\enddata
\tablenum{1}
\end{deluxetable}

\newpage
\begin{deluxetable}{llllll}
\tablecaption{(continued)}
\tablehead{
\colhead{SN} & \colhead{Galaxy} & \colhead{DDO Type} &
\colhead{SN Type} & \colhead{Velocity (km/s)} & \colhead{Remarks}
}
\startdata
2002dj &   N5018     & E2/Sa pec        &   Ia   &    2794 & \\
2002dk &   N6616     & Sa t             &   Ia-pec (91bg)  &    5556    &   2 \\
2002dl &   U11994    & S IV             &   Ia-pec (91bg)  &   4872     &   2 \\
2002dn &   I5145     & Sab              &   Ic   &    7355& \\
2002do &   M+07-41-1 & E2               &   Ia-pec (91bg)  &    4761& \\
2002dp &   N7678     & Sc pec II:       &   Ia   &    3489& \\
2002dq &   N7051     & S(B?)ab II       &   II   &    2519& \\
2002dr &   U12214    & E3               &   Ia   &    6610& \\
2002ds &   ESO 581-G25&S IV             &   II   &    2277    &   2 \\
2002dt &   ESO 516-G5& Sc I-II          &   Ib/c? &    7487& \\
2002dv &   U11486    & S pec            &   II   &    7923& \\
2002dw &   U11376    & Merger           &   II   &    6528& \\
2002dx &   U12861    & Sa               &   Ia   &    7100& \\
2002dy &   M-01-59-24& Sc               &   II   &    9904& \\
2002dz &   M-01-1-52 & S III-IV:        &   Ib/c  &    5361& \\
2002ea &   N 820     & Sa               &   IIn  &    4418& \\
2002eb &   CGCG 473-011& Sa:              &   Ia   &    8255& \\
2002ec &   N5910     & E0 (t?)          &   Ia   &   11976& \\
2002ed &   N5468     & Sc I-II          &   II   &    2845& \\
2002ee &   N5772     & Sa               &   II   &    4900& \\
2002ef &   N7761     & E0               &   Ia   &    7080& \\
2002eg &   U11486    & S pec            &   IIb  &    7923& \\
2002eh &   N 917     & Sa               &   Ia   &    5388& \\
2002ei &   M-01-09-024&S IV             &   II   &    2319& \\
2002el &   N6986     & E3               &   Ia   &    9000& \\
2002em &   U3430     & Sab:             &   II   &    4059    &   2 \\
2002en &   U12289    & S                &   II   &   10163& \\
2002eo &   N 710     & Sc: III-IV:      &   II   &    6132& \\
2002er &   U10743    & Sab III-IV:      &   Ia   &    2569    &   2 \\
2002es &   U2708     & E1/Sa            &   Ia-pec (91bg)  &    5394& \\
2002et &   M-04-47-10& Sb II            &   Ia   &    8217& \\
2002eu &     ...     & S0/Sa            &   Ia-pec (91bg)  &   11280& \\
2002ey &     ...     & E2/S0            &   Ia-pec (91bg)  &     ...& \\
2002fb &   N 759     & E1               &   Ia-pec (91bg)  &    4667& \\
2002fi &   M-04-7-10 & SBb II-III       &   Ia   &   17138& \\
2002fj &   N2642     & SBbc I           &   IIn  &    4342& \\
2002gc &   U1394     & Sb III-IV:       &   Ia-pec (91bg)  &    6388& \\
2002gd &   N7537     & Sab III:         &   II   &    2674& \\
2002gw &   N 922     & S(B)c: III:      &   II   &    3092& \\
2002gy &   U2701     & Sbc IV           &   Ib/c  &    7285    &   2 \\
2002ha &   N6962     & Sbc II           &   Ia   &    4211& \\
\enddata
\tablenum{1}
\end{deluxetable}

\newpage
\begin{deluxetable}{llllll}
\tablecaption{(continued)}
\tablehead{
\colhead{SN} & \colhead{Galaxy} & \colhead{DDO Type} &
\colhead{SN Type} & \colhead{Velocity (km/s)} & \colhead{Remarks}
}
\startdata
2002hc &   N2559     & Sc: pec          &   II   &    1561& \\
2002hd &   M-01-23-8 & E3               &   Ia   &   10500& \\
2002he &   U4322     & E2               &   Ia-pec (91bg)  &    7384& \\
2002hf &   M-05-3-20 & Sab II:          &   Ic   &    5609& \\
2002hg &   N3306     & Sb: III-IV:      &   II   &    2889& \\
2002hh &   N6946     & Sc I             &   II   &      48& \\
2002hi &     ...     & ?                &   IIn  &   18000    &   7 \\
2002hk &   M-07-15-6 & S(B:)ab III      &   II   &    5751& \\
2002hl &   N3665     & E1               &   Ia   &    2080& \\
2002hm &   N4016     & S(B?) pec        &   II   &    3448& \\
2002hn &   N2532     & Sc II-III:       &   Ic   &    5260& \\
2002ho &   N4210     & SBbc II          &   Ic   &    2732& \\
2002hv &   U4974     & E0               &   Ia   &    7023& \\
2002hw &   U 52      & Sa               &   Ia   &    5257& \\
2002hx &     ...     & SBbc II          &   II   &    9293& \\
2002hy &   N3464     & Sb II            &   Ib   &    3729& \\
2002hz &   U12044    & Sb/S0            &   Ib/c  &    5444& \\
2002je &     ...     & S pec            &   II   &     ...& \\
2002jg &   N7253     & Ir/Pec t         &   Ia   &    4718& \\
2002ji &   N3655     & S pec            &   Ib/c  &    1473& \\
2002jj &   I 340     & S pec?           &   Ic   &    4218    & 10  \\
2002jm &   I 603     & Sa               &   Ia-pec (91bg)  &    5400    &  5 \\
2002jo &   N5708     & S pec            &   Ia   &    2751& \\
2002jp &   N3313     & SBb I            &   Ic   &    3706& \\
2002jy &   N 477     & S(B)c II:        &   Ia   &    5876& \\
2002jz &   U2984     & Pec              &   Ic   &    1543& \\
2002ka &     ...     & E1/S0            &   Ia   &    2062		& \\
2003A &    U5904     & Sb               &   Ib/c  &    6591   &   2,8 \\
2003C &    U 439     & Sa pec           &   II   &    5302& \\
2003D &    M-01-25-9 & E3               &   Ia-pec (91bg)  &    6628    &   6 \\
2003E &    M-04-12-4 & Sc               &   II   &    4409    &   2 \\
2003F &    U3261     & Sc               &   Ia   &    5169& \\
2003G &    I 208     & Sa               &   IIn  &    3449 & \\
2003H &    N2207     & Sc I-II: t + St  &   Ib   &    2741    &   9 \\
2003I &    I2481     & Sab              &   Ib   &    5322 & \\
2003J &    N4157     & Sb III           &   II   &     774    &   2 \\
2003K &    I1129     & Sc               &   Ia   &    6540& \\
2003L &    N3506     & Sc III:          &   Ic   &    6403& \\
2003M &    U7224     & ?                &   Ia   &    7267    &   7 \\
2003O &    U2798     & S(B)b III-IV     &   II   &    4941& \\
2003S &    M+09-22-94& S                &   Ia   &   11700    &   2 \\
\enddata
\tablenum{1}
\end{deluxetable}
\newpage
\begin{deluxetable}{llllll}
\tablecaption{(continued)}
\tablehead{
\colhead{SN} & \colhead{Galaxy} & \colhead{DDO Type} &
\colhead{SN Type} & \colhead{Velocity (km/s)} & \colhead{Remarks}
}
\startdata
2003T &    U4864     & S pec            &   II   &    8268& \\
2003U &    N6365     & Sbn:             &   Ia   &    8496& \\
2003W &    U5234     & Sc III:          &   Ia   &    6017& \\
2003X &    U11151    & S0:              &   Ia   &    7017& \\
2003Y &    I 522     & E1               &   Ia-pec (91bg)  &    5079& \\
2003Z &    N2742     & Sc II-III        &   II   &    1289& \\
2003aa &   N3367     & S(B)c I-II       &   Ia-pec (91T)  &    3037& \\
2003ac &   I3203     & S0:              &   IIb  &    6910    &   2 \\
2003ag &   U6440     & Sc I             &   Ia   &    6777 & \\
2003ah &     ...     & Sb:              &   Ia   &     ...    &   2 \\
2003ai &   I4062     & Sa?              &   Ia   &     ...& \\
2003am &   ESO 576-G40&S0: t            &   II   &    2085& \\
2003an &   M+05-32-22& E1               &   Ia   &   11163& \\
2003ao &   N2993     & Sn t             &   II   &    2420 & \\
2003au &   N6095     & E0               &   Ia-pec (91bg)  &    9243& \\
2003bk &   N4316     & Sab III:         &   II   &    1252    &   2 \\
2003bl &   N5374     & Sc II:           &   II   &    4295& \\
2003bm &   U4226     & S pec            &   Ic   &    7907& \\
2003bp &   N2596     & Sb:              &   Ib/c  &    5938& \\
2003bq &   U3513     & Sb III-IV:       &   ?    &    7358& \\
2003br &   M-05-34-18& Sb I-II          &   II   &    3758& \\
2003bu &   N5393     & Sa:              &   Ic   &    6019& \\
2003bw &   I1077     & Sc II:           &   II   &    3452& \\
2003cg &   N3169     & Sab III          &   Ia   &    1238& \\
2003ch &   U3787     & E3               &   Ia   &     ...& \\
2003ci &   U6212     & Sb: II: t        &   II   &    9091& \\
2003cm &   U10590    & SBbc III:        &   ...  &    3054& \\
2003cn &   I 849     & Sab:             &   II   &    5430& \\
2003cq &   N3978     & Sbn: pec         &   Ia   &    9978& \\
2003dg &   U6934     & Sb: III-IV       &   ...  &    5536    &   2 \\
2003dl &   N5789     & S*/Ir            &   ...  &    1805& \\
2003dr &   N5714     & Sbc:             &   ...  &    2237    &   2 \\
2003ds &   M+08-19-17& ?                &   ...  &    9210 & \\
2003dt &   N6962     & Sb II t?         &   Ia   &    4211& \\
2003du &   U9391     & SB/Ir IV:        &   Ia   &    1914& \\
2003dv &   U9638     & Sc III-IV:       &   IIn  &    2271& \\
2003dw &   M+10-24-51& Sb II            &   Ia   &    9003& \\
2003ed &   N5303     & S?               &   IIb  &    1419    &   4  \\
2003ef &   N4708     & S pec            &   II   &    4103& \\
\enddata
\end{deluxetable}
\newpage

\begin{deluxetable}{llllll}
\tablecaption{Massive Supernovae in Possible E and S0 Galaxies}
\tablehead{
\colhead{SN} & \colhead{Galaxy} & \colhead{DDO Type} &
\colhead{SN Type} & \colhead{Velocity (km/s)} & \colhead{Remarks}
}
\startdata 
2000ds   & N2768  & E3/Sa &Ib/c  &1373 &  1\\
2001I    & U2836  & E2    &IIn  &4963 &  2\\
2002at   & N3720  & E1    &II   &5985 &  3\\
2002bx   & I2461  & S0    &II   &2260 &  4\\
2003ac   & I3203  & S0:   &II   &6910 &  5\\
\enddata
\tablenotetext{1} {On the basis of a plate obtained by 
  Walter Baade with the Palomar 200-in telescope, Sandage \& Bedke 
  (1994) state that ``The definite outer envelope of NGC 2768
   surrounding an E6 bulge makes the S0 classification certain."}
\tablenotetext{2} {The presence of this SN~IIn in an E
   galaxy was already noted in Paper I. However, inspection
   of images at different contrast shows evidence for
   some substructure in the envelope of this galaxy. An
   Sa: classification might therefore be more appropriate
   than the one given in the tables.}
\tablenotetext{3} {Inspection of panel 185 of the Carnegie Atlas of Galaxies
   (Sandage \& Bedke 1994) appears to show an image of NGC 3719,
   rather than that of NGC 3720, which  was the host galaxy 
   of SN 2002at. The morphological classification of NGC 3720 
   given in A Revised Shapley Ames Catalog of Bright Galaxies 
   (Sandage \& Tammann 1981) also appears to refer to NGC 3719.
   A. Sandage (2003, private communication) informs us
   us that such an interchange would not have 
   affected the classification given in the Revised Shapley-Ames 
   Catalog because both NGC 3719 and NGC 3720 are assigned to the 
   same classification type. However, our inspection of red
   and blue images with differing contrast actually shows some
   structure in the envelope of NGC 3720, suggesting that an Sa: classification
   would be preferable to the the E1 classification given in our
   tables.}
\tablenotetext{4} {The classification is somewhat uncertain because this
object is viewed edge-on. It might be an edge-on S0 galaxy that contains 
   a dust lane. A blue image shows some clumpiness in the disk,
   suggesting the presence of a young population component.}
\tablenotetext{5} {The classification is somewhat uncertain because this 
   object is viewed edge-on. Reexamination of this object on images
   with differing contrast suggests some distortions of the
   structure. Perhaps S0-pec might therefore
   have been a more appropriate classification than the S0: 
   classification given in our tables. }
\end{deluxetable}

\newpage

\begin{deluxetable}{llllll}
\tablecaption{Galaxy Classification and SN Type: New\tablenotemark{a,b}}
\tablehead{
\colhead{Galaxy type} & \colhead{Ia} & \colhead{Ia-pec} &
\colhead{Ibc$^c$} & \colhead{II} & \colhead{IIn}
} 
\startdata
E         &   14.5 & 9.5  & 0  &  1 &   0\\
E/Sa      &    3   & 1    & 1  &  0 &   0\\
Sa        &    9   & 4    & 2  &  2 &   2\\
Sab       &    5   & 0    & 2  &  9 &   0\\
Sb        &   16.5 & 1    & 7.5& 19 &   2\\
Sbc       &    6   & 2    & 7  &  6 &   1\\
Sc        &   11   & 1    & 8  & 22 &   2\\
Ir        &    2   & 0    & 0  &  0 &  0.5\\
\enddata
\tablenotetext{a} {Only new data from Table 1.}
\tablenotetext{b} {Half-integer values refer to intermediate 
morphologies (e.g., E/Sb is counted as 0.5 E and 0.5 Sb).}
\tablenotetext{c} {Here, the ``Ibc" designation includes
SNe~Ib, Ib/c, and Ic.} 
\end{deluxetable}

\begin{deluxetable}{llllll}
\tablecaption{Galaxy Classification and SN Type: All\tablenotemark{a,b}}
\tablehead{
\colhead{Galaxy type} & \colhead{Ia} & \colhead{Ia-pec} &
\colhead{Ibc$^c$} & \colhead{II} & \colhead{IIn}
}
\startdata
E         &   21.5 & 10.5 &  0  &  2 &  1\\
E/Sa      &    8   &  3   &  1  &  0 &  0\\
Sa        &   13   &  5   &  4  & 10 &  2\\
Sab       &    9   &  4   &  4  & 11 &  0\\
Sb        &   35.5 &  3   &  9.5& 36 &  4\\
Sbc       &   11   &  3   & 13  & 18 &  2\\
Sc        &   17   &  1   & 15  & 40 &  6\\
Ir        &    2   &  0   &  0  &  2 &  0.5\\
\enddata
\tablenotetext{a}{All host galaxies of SNe discovered during LOSS/LOTOSS --
i.e.,  the sum of the data from the present paper and those from Paper I.}
\tablenotetext{b} {Half-integer values refer to intermediate 
morphologies (e.g., E/Sb is counted as 0.5 E and 0.5 Sb).}
\tablenotetext{c} {Here, the ``Ibc" designation includes
SNe~Ib, Ib/c, and Ic.} 
\end{deluxetable}

\begin{deluxetable}{lll}
\tablecaption{Frequency of Host Galaxies of Normal SNe~Ia and SNe~Ia-pec\tablenotemark{a}}
\tablehead{
\colhead{Host Galaxy Type} & \colhead{Normal SNe~Ia} & \colhead{SNe~Ia-pec}
}
\startdata 
E + S0 Galaxies     &      26.5  &        14\\
Spirals + Irregulars &     101    &        20\\
\enddata
\tablenotetext{a} {Half-integer values refer to intermediate 
morphologies (e.g., E/Sb is counted as 0.5 E and 0.5 Sb).}
\end{deluxetable}

\end{document}